\documentclass{aipproc}

\usepackage{graphics}
\usepackage{color}

\usepackage{longtable}

\begin{document}

\title{COSMIC RAY ACCELERATION IN SUPERBUBBLES
AND THE COMPOSITION OF COSMIC RAYS}
\author{R.E. Lingenfelter$^1$, J.C. Higdon$^{2}$, \& R. Ramaty$^3$}
\affiliation{1) Center for Astrophysics and Space Sciences, University of
California, San Diego, La Jolla, CA \\ 2) W. M. Keck Science Center, Claremont
Colleges, Claremont, CA \\ 3) Laboratory for High Energy Astrophysics,
NASA Goddard Space Flight Center, Greenbelt, MD}

\begin{abstract}
We review the evidence for cosmic ray acceleration in
the superbubble/hot phase of the interstellar medium, and discuss the
implications for the composition of cosmic rays and the structure and
evolution of the interstellar medium (ISM).
We show that the bulk of the galactic supernovae, their expanding
remnants, together with their metal-rich grain and gas ejecta, and
their cosmic ray accelerating shocks, are all confined within the
interiors of hot, low-density superbubbles, generated by the multiple
supernova explosions of massive stars formed in giant OB associations.
This superbubble/hot phase of the ISM provides throughout the age of
the Galaxy a cosmic ray source of essentially constant metallicity for
acceleration by the shocks of many supernovae over time scales of a
few Myr, consistent with both Be/Fe evolution and ACE observations of $^{59}$Ni/$^{59}$Co.
We show that the expected metallicity ($>$ 2 times Solar) and
filling factor ($>$ 50\%) of the superbubble/hot phase is high enough
that the composition of cosmic rays accelerated from fast,
supernova grains in these superbubbles is quite consistent with
the both Be/Fe and cosmic ray data, while their acceleration from grains
in the well-mixed cooler phases of the ISM is {\em not} consistent
with observations. We also show that if the refractory cosmic ray
metals come from the sputtering of fast refractory grains then the
accompanying scattering of ambient gas by these fast grains can also
account for the relative abundance of cosmic ray volatiles.

\end{abstract}

\maketitle

\section{Introduction}
%\label{intro.sec}

ACE measurements by Wiedenbeck et al. (53) of the cosmic ray $^{59}$Ni/$^{59}$Co
abundance ratio and optical measurements of the Be/Fe abundance
ratios in old stars, e.g. Molaro et al. (34) and Boesgaard et al. (4),
have shown us that the making of cosmic rays is like the making of fine
wine. Both have to be aged and blended. Cosmic rays can not be too young
and fresh, because their low ratio of K-capture $^{59}$Ni to its daughter
$^{59}$Co requires an age of $> 10^5$ years before acceleration, and they
can not be too old and diluted, because the roughly constant ratio of
cosmic ray produced Be to supernova produced Fe requires the metallicity
in the matter from which the cosmic rays are accelerated to be roughly
constant within a factor of $\sim$2. This can be achieved if the
metallicity is dominated by supernova ejecta.

The cellar where all of this aging and blending happens quite naturally
is the vast metal-enriched superbubble/hot phase of the interstellar
medium (SB/HISM), where most supernovae occur. This can be seen from
the extensive studies of Galactic and extragalactic supernovae, their
progenitors and their cosmic ray accelerating remnants, and of the
chemical enrichment and evolution of the interstellar medium, that
we will review here with special emphasis on the metallicity and filling
factor of the SB/HISM. We also discuss how ACE and other measurements
of the cosmic ray composition can define these two properties
in the region of cosmic ray acceleration.

The source of energy for cosmic ray acceleration is thought to be
shock waves driven by the expansion energy of supernova ejecta,
e.g. Blandford \& Ostriker (3) and Axford (1). The power required
to maintain the Galactic cosmic rays is about 10$^{41}$ ergs$^{-1}$,
e.g. Lingenfelter (23). The Galactic supernova rate is about 3
supernovae per century, e.g. van den Bergh \& McClure (49), most of
which (80 to 90\%) are core-collapse (Type II and Ib/c) supernovae
of relatively young ($<$ few 10$^7$ yrs) massive O and B stars; the
remainder are Type Ia thermonuclear explosions of much older accreting
white dwarfs. Thus, the average cosmic ray energy needed per supernova
is about 10$^{50}$ ergs, which requires an acceleration efficiency
of about 10\% from the blast wave shocks of the supernovae, since
supernovae all seem to have similar ejecta kinetic energies of about
10$^{51}$ ergs, e.g. Woosley \& Weaver (56) and Nomoto et al. (36).

The source of the particles that are accelerated as cosmic rays and the
site of their acceleration, however, are still debated. But if the energy
comes from supernova shocks, the site and source of the particles clearly
must be in the material through which the shocks pass. New clues to origin
of the particles come from recent measurements, e.g. Molaro et al. (34) and
Boesgaard et al. (4), of Be/Fe abundances in old halo stars that show
that the ratio of cosmic ray spallation produced Be relative to core-collapse
supernova-produced Fe has remained roughly constant throughout the evolution
of the Galaxy. This constancy requires (Ramaty, Kozlovsky \& Lingenfelter (38);
Ramaty, Lingenfelter \& Kozlovsky (40)) that the cosmic rays be accelerated
out of matter that is only partially diluted by mixing with the interstellar
medium (ISM) so that it is still sufficiently enriched in supernova-synthesized
metals that its metallicity did not changed by more than $\sim$2 over galactic
evolution. Detailed calculations by Ramaty et al.
(42) and Ramaty, Lingenfelter \& Kozlovsky (41) of the production and
evolution of the Galactic Be/Fe ratio clearly show that the bulk of the
cosmic rays can not be accelerated from the well-mixed ISM, as has been
recently assumed, e.g. Meyer et al. (33) and Ellison et al. (11).

Just such a metal-enriched environment is where most supernovae do in fact
occur and through which most of the cosmic ray accelerating, supernova
shocks propagate. As we have discussed  in Higdon, Lingenfelter \& Ramaty
(17,18), extensive observations show that the bulk of the core-collapse
supernova progenitors are formed in OB associations in giant molecular
clouds and that the combined winds and supernova ejecta of these stars
form hot, low density superbubbles, that reach dimensions of several
hundred pc and last for tens of Myr. During this time the bulk of the
supernova ejecta, the supernova shocks, and their cosmic ray acceleration
are all confined within the superbubble/hot phase of the ISM.

\section{Supernovae \& the Superbubble/Hot ISM}
%\label{origin.sec}

Core-collapse (Type II and Ib/c) supernovae are highly correlated
in space and time, e.g. McCray \& Snow (28). Such supernovae thus
create giant cavities, or superbubbles, in the interstellar medium
rather than many smaller, isolated bubbles, e.g. Mac Low \& McCray
(26) and Tomisaka (48). This is expected because 1) the massive O and B
star supernova progenitors ($>$8 M$_{\odot}$; e.g. Woosley \& Weaver
(56); Nomoto et al. (36)) are not distributed uniformly in interstellar
space, but tend to form clusters, since the majority of these massive
stars are born in the most massive ($>$ 10$^{5}$ M$_{\odot}$) molecular
clouds in gravitationally unbound OB associations, while less massive
clouds are destroyed by the intense UV irradiation with the birth
of their first O star (McKee \& Williams (30)); and 2) these stars
are short-lived and slow moving; the progenitors of core-collapse
supernovae have main sequence lifetimes of $\sim$ 3 to 35 Myr and
OB stars in associations have dispersion velocities of only $\sim$4
km s$^{-1}$ (Blaauw (2)), so they do not travel too far ($\sim$120
pc in 30 Myr) from their birthplaces before they die in supernova
explosions. Consequently, the combined effect of these clustered
supernova explosions is to create superbubbles, which expand and
merge to form the hot ($>$10$^6$ K), tenuous ($<$10$^{-3}$ cm$^{-3}$)
phase of the ISM with an average filling factor of $\sim$ 50\%, or
more (e.g. Yorke (57); Spitzer (45); McKee (29); Rosen \& Bregman
(43); Korpi et al. (20)) of the Galactic disk (with a scale height
$\sim$100 pc) and essentially all of the corona/halo (with scale
height of $\sim$3 kpc). We discuss the spatial distribution of the
SB/HISM filling factor in more detail below.

An analysis of the surface brightness distribution of the remnants of
historical supernovae in our Galaxy by (Higdon \& Lingenfelter (16)
has shown that 85$\pm$10\%, of the observed Galactic supernovae occured
in the superbubble hot phase of the ISM. This is quite consistent with
more extensive observations of supernovae in other late type galaxies.
As discussed in detail in Higdon et al. (17), the combined observations
of van Dyk et al. (51) and Kennicutt, Edgar, \& Hodge (19) show that
the great majority, $\sim$ 90$\pm$10\%, of the core-collapse supernovae
in late type galaxies also occur within superbubbles, and because of the
large filling factor of the superbubble, hot-phase of the ISM, half, or
more, of the Type Ia should also occur within the superbubbles just by
chance. Thus, since core-collapse supernovae account for 80 to 90\% of
all supernovae in our galaxy and Type Ia make up the remainder, roughly
80\% of all supernovae occur in the superbubble hot phase of the ISM,
and the bulk of the cosmic rays accelerated by their shocks are also
produced there.

The observed concentration of supernovae in the SB/HISM and the subsequent
cosmic ray acceleration in this hot ($>10^6$ K) phase also argues strongly
against a first-ionization-potential (FIP) injection bias for cosmic
ray enrichment which requires warm partially ionized gas,
and does not work in the nearly fully ionized gas of the SB/HISM. However,
the SB/HISM environment is quite consistent with cosmic ray source particle
injection from the sputtering of the high velocity refractory grains formed
in supernovae, see Lingenfelter, Ramaty \& Kozlovsky (25) and Lingenfelter \&
Ramaty (24). Such a volatility bias for cosmic ray refractory metal injection
from sputtering of supernova grains was first suggested by Cesarsky \& Bibring
(7) and it was also recently proposed  by Meyer et al. (33) for shock-accelerated,
ice-stripped, refractory cores of grains in the warm ISM.

The occurrence of most supernovae in the SB/HISM and cosmic ray acceleration
in that hot phase also argues strongly against the mass/charge (A/Q)-dependent
acceleration model of Ellison et al. (11) for the volatile elements, because
this too requires warm partially ionized gas, and Ellison \& Meyer (12) argue
it does not work in the highly ionized gas of the SB/HISM. As we have shown
in Lingenfelter \& Ramaty (24) and discuss below, however, a mass dependent
injection of volatiles appears to result quite naturally from the scattering
of the ambient gas atoms in direct collisions with fast grain atoms that must
accompany the sputtering of the grains.

\section{Metallicity of the Superbubble/Hot ISM}
%\label{metallicity.sec}

The bulk of the metals (elements with Z$>$5) in the Galaxy have been
produced by supernovae and ejected into the ISM. The relative abundances
of most elements have remained relatively constant (e.g. Timmes, Woosley
\& Weaver (47)), because they simply reflect the IMF-averaged supernova
yields which do not depend strongly on the interstellar metallicity,
e.g. Woosley \& Weaver (56). The averaged relative abundances of the
present ISM, e.g. Savage \& Sembach (44), the older (4.5 Gyr) Solar
system material, e.g. Grevesse, Noels \& Sauval (15), and the
IMF-averaged fresh supernova ejecta, e.g. Lingenfelter et al. (25), are
all within about $\sim$ 10\% of one another. But their overall abundance
in the ISM (i.e. the interstellar metallicity) has grown steadily over
time with the accumulation of fresh supernova ejecta continuously
injected and mixed into the ISM. The time scale for thorough mixing
is generally thought (e.g. McWilliam (31); Thomas, Greggio \& Bender
(46)) to be on the order of 30 to 100 Myr. This is comparable to the
typical mean life of the SB/HISM reservoir into which the bulk of the
supernova ejecta with a metallicity 10 times Solar (e.g. Woosley \&
Weaver (56)) is injected and in which the bulk of the mixing is expected
to occur. Thus, we would expect significant variations in the metallicity,
but not in the abundances of most elements relative to one another, within
the SB/HISM as a function of the age of individual superbubbles and their
generating OB associations.

The average, or equillibrium, metallicity of the SB/HISM is not known,
but the supernova ejecta appear to be able to provide sufficient metals
to produce a metallicity $>$2 times Solar, as is required (Ramaty et al.
2000b) for the cosmic ray source from the constancy of the Be/Fe abindances
in old stars. This can be seen from a simple comparison of the total mass
of the SB/HISM and the mass of supernova ejecta produced during the mean life
of the SB/HISM. The total mass of the SB/HISM is $\sim 10^8$ M$_{\odot}$,
assuming an average, e.g. Yorke (57) and Spitzer (45), SB/HISM density of
$\sim 10^{-3}$ H/cm$^3$, a SB/HISM
scale height of $\sim$ 3 kpc and an effective Galactic radius of $\sim$ 15
kpc. Taking a nominal SB/HISM mean life, or mixing time, $t \sim$ 100
Myr, the required SB/HISM input is $\sim 1$ M$_{\odot}$/yr($t$/100Myr).
The present Galactic SNII/Ibc rate of about 1 SN every 40 yr, producing
an IMF-averaged ejecta mass of 18 M$_{\odot}$, gives a Galactic SNII/Ibc
ejecta input of $\sim$ 0.45 M$_{\odot}$/yr with a metallicity, $z_{SN}$
of 10 times Solar. If all of the remaining SB/HISM mass comes from
evaporated clouds and swept up gas in the well-mixed ISM with Solar
metallicity $z_{\odot}$ of 1, then the averaged SB/HISM metallicity,
$z_{HISM} \sim [10M_{SN} + 1(M_{HISM} - M_{SN})]/M_{HISM}$, or
$z_{HISM} \sim 1 + 9M_{SN}/M_{HISM} \sim$ 1 + 4($t$/100Myr).
Thus the SB/HISM metallicity $z_{HISM} >$ 2 times Solar for any SB/HISM
mean life, or mixing time, $t >$ 25 Myr, consistent with the estimated
values, e.g. McWilliam (31) and Thomas, Greggio \& Bender (46).

Such a mixing time, or mean age, of metals from supernova ejecta in
these superbubbles is more than a couple orders of magnitude longer
than the minimum age ($<$100 kyr) of cosmic ray source metals required
by the ACE observations of Wiedenbeck et al. (53), showing that the bulk
of the $^{59}$Ni had decayed (with a 110 kyr mean life) in the cosmic
ray source material prior to acceleration.

Observational evidence of such supernova ejecta enriched superbubble
metallicity may be found in the x-ray emission from the interiors of
giant HII regions in the Large Magellanic Cloud (LMC), thought to be
superbubbles powered by supernovae. The observed x-ray luminosities
of these bubbles, which should scale directly with metallicity, are
an order of magnitude higher than would be expected (Chu \& Mac Low
(8)) if they had a typical LMC metallicity of only 1/3 Solar. Thus,
we suggest that the x-ray observations do in fact imply an average
metallicity of roughly 3 times Solar in these superbubbles.

Such metallicities are larger than that calculated from the simple
analytic superbubble model of Mac Low \& McCray (26), which
assumes conductive heating and evaporation of swept-up ISM as
the primary source of superbubble gas and predicts an averaged
metallicity of only $\sim$ 1.1 for a $\sim$ 50 Myr old superbubble
of $\sim$ 700 pc radius. But this model was based on several
assumptions that greatly reduce the superbubble metallicity.
First, the model neglected the interstellar magnetic fields that
would greatly supress the conductive heating normal to the field
lines, overestimating the ISM input, and also provide additional
external confining pressure, underestimating the supernova input
required to generate the bubble, e.g. Tomisaka (48). In addition,
the model assumed the unit density ISM extended to heights much
larger than the 700 pc bubble radius, instead of the measured
scale height of $<$ 200 pc, which would greatly reduce the assumed
ISM input, and moreover would allow the blow-out of the superbubble
into the Galactic halo, also greatly underestimating the supernova
input required to generate the bubble. As a result of these
underestimates of the required supernova power, the model was able
to generate a $\sim$ 50 Myr old superbubble of $\sim$ 700 pc radius,
with a very low effective rate of SNII/Ibc supernovae of only $\sim$
2 SN/Myr kpc$^2$ in the Galactic plane. This assumed rate is only 3\%
of the estimated local SNII/Ibc rate $\sim$ 70 SN/Myr kpc$^2$ (as we
show below), and thus the model underestimates the supernova ejecta
input by a factor of 35! When an appropriate supernova rate is used
and even minimal effects of the magnetic fields and gas scale height
are considered, superbubble metallicities of more that 2 times Solar
would be expected.

A large fraction of the C, O and refractory metals in this ejecta may be
in graphite and oxide grains, since in the core-collapse supernova 1987A
roughly 0.2 M$_{\odot}$ of this material condensed out of the cooling,
expanding ejecta as high velocity ($\sim$ 2500 km/s) grains within 2 years
after the explosion, see Kozasa, Hasegawa \& Nomoto (21), and as much as 1
M$_{\odot}$ could be expected, see Dwek (9), to condense before the ejecta
is reheated and slowed by the reverse shock and the grains with a much
smaller charge to mass ratio begin to move separately from the ejecta
plasma. In fact, Dwek (9,10) suggests that supernova ejecta are the
major source of refractory grains in the Galaxy and interactions with
supernova shocks are the major cause of their destruction.

Thus, supernova ejecta and winds can be expected to dominate the metallicity
and grains within the SB/HISM, where the bulk of supernova shock waves
are dissipated and the bulk of cosmic rays should be accelerated. These
supernova grains should therefore be the major injection source required for
the cosmic ray metals, because of their high initial velocity (Lingenfelter
et al. (25)) and possible subsequent acceleration (Ellison et al. (11)).
Moreover, because the metallicity of the supernova ejecta is essentially
independent of progenitor metallicity (Woosley \& Weaver (56)), the SB/HISM
can provide the essentially constant source of cosmic ray metals required
by Be/Fe observations. Therefore, we would expect that throughout the age
of the Galaxy, the bulk of the core-collapse supernovae occur in the metal
enriched SB/HISM, and the blast wave shocks of their remnants accelerate the
bulk of the Galactic cosmic rays out of the enriched gas and dust in the SB/HISM.

\section{Filling Factor of the Superbubble/Hot ISM}
%\label{fillingfactor.sec}

The hot ($\sim 10^6$ K), tenuous ($\sim 10^{-3}$ H/cm$^3$) phase of
the interstellar medium is powered primarily by Galactic supernovae
and formed through the merger of superbubbles, generated by the
clustered supernovae in OB associations. This can be seen energeticly
from a comparison of the power required to maintain the pressure in
the SB/HISM and that provided by Galactic supernovae, which suggests
that the filling factor, i.e. the fractional volume, of the SB/HISM
should be large. The total energy in SB/HISM is $\sim 3\times10^{56}
f_{HISM}$ ergs, assuming a SB/HISM filling factor, $f_{HISM}$, a
SB/HISM pressure of $\sim 3\times10^{-12}$ erg/cm$^3$, a Galactic
radius of 15 kpc and a scale height of 3 kpc. For a SB/HISM mean life
$t$ of 100 Myr, the power required to maintain the SB/HISM is $\sim
3\times10^{48} f_{HISM}$ ergs/yr($t$/100Myr). The Galactic SNII/Ibc
rate of $\sim$ 1 SN/40 yr with an average ejecta energy $\sim 10^{51}$
ergs/SN, gives a Galactic SNII/Ibc power of $\sim 2.5\times10^{49}$
ergs/yr. Thus even with significant ($>$50\%) energy losses SNII/Ibc
could completely fill ($f_{HISM}$ = 1) the Galaxy with the SB/HISM in
$t >$ 25 Myr.

The overall Galactic average value of the filling factor of SB/HISM
is, in fact, generally taken to be $\sim$ 50\%, or more, depending on
the assumed Galactic scale height, e.g. Yorke (57), Spitzer (45),
McKee (29). Because of the strong dependence of the SB/HISM filling
factor on local supernova rates, it is thought to be high $\sim$
90\% in the inner Galaxy (i.e. within the Solar radius of 8.5 kpc)
where most Galactic supernovae occur, as well as in the Galactic halo
where the superbubbles blow-out, and low $<$ 50\% in the outer Galaxy
beyond the Solar radius where few supernovae occur.

A more quatitative estimate of the dependence of the SB/HISM filling
factor on Galactic radius (see Table~\ref{tab:a}) can be made from
the radial dependence of the Galactic supernova rate and the calculated
filling factor versus the supernova rate.

\begin{table}
\begin{tabular}{ccccc}
\hline
   Galactic  & MoleCloud   & OBAssoc     & Supernova      & SB/HISM \\
    Radius   & Density     & Density     & Rate
\tablenote{Galactic SN rate of 3 SN/100yr normalized to surface density
distribution of molecular clouds from Williams \& McKee (54) and OB
associations from McKee \& Williams (30)}
      & Filling Factor\tablenote{Expected filling
factor based on the SN rate from the hydrodynamic calculations
by Rosen \& Bregman (43) and Korpi et al. (20).}  \\
      kpc  & M$\odot$/pc$^2$ & N/kpc$^2$ & SN/kpc$^2$Myr & $\mid z\mid<$300pc \\
\hline
        1    &    1     &      0.3      &   35     &     $\sim$ 0.4  \\
        4    &    7     &      2.3      &  250     &     $\sim$ 0.9  \\
        6    &    6     &      1.6      &  210     &     $\sim$ 0.9  \\
        8    &   2.5    &      0.6      &   90     &     $\sim$ 0.9  \\
       10    &   1.5    &      0.4      &   50     &     $\sim$ 0.5  \\
       15    &   0.4    &       -       &   12     &     $\sim$ 0.1  \\
\hline
\end{tabular}
\caption{RADIAL DEPENDENCE OF GALACTIC SUPERNOVA RATE \& EXPECTED
SB/HISM FILLING FACTOR}
%Radial Dependence of the Galactic Supernova Rate and the Expected
%SB/HISM Filling Factor}
\label{tab:a}
\end{table}

The dependence of the SB/HISM filling factor on local supernova rates
has been quantified by recent calculations of 2D hydrodynamics by Rosen
\& Bregman (43), and 3D magnetohydrodynamics by Korpi et al. (20).
These calculations determined the filling factors of all phases of the
ISM as a function of height $z$ above the Galactic plane for a range
of supernova rates. Generally these calculations suggest that the
SB/HISM filling factor is lowest at the Galactic plane where the
superbubble expansion is most constrained by the warm and cold phases
of the ISM, and increases to $\sim$ 100\% in the halo at large distances
above the plane. For the purposes of cosmic ray acceleration and
composition, what is important is the height averaged filling factor
of the SB/HISM for $\mid z \mid <$ 300 pc, which is the range of heights where
most of the supernovae occur. For assumed supernova rates (adjusted
to $10^{51}$ ergs/SN) of 5, 20, 40 and 80 SN/Myr kpc$^2$ in the Galactic
plane these calculations give height averaged ($\mid z \mid <$300 pc)
SB/HISM filling factors of $\sim$ 0.1, 0.4, 0.6 and 0.9 respectively.
This suggests that at low supernova rates ($<$ 40 SN/Myr kpc$^2$) the
SB/HISM filling factors within $\mid z \mid <$300 pc scale roughly linearly
with the supernova power, while at higher supernova rates ($\geq$ 80 SN/Myr
kpc$^2$) the SB/HISM filling factors within $\mid z \mid <$300 pc reach
a maximum value of $\sim$ 90\%.

The Galactic radial dependence of the supernova rate can be estimated by
normalizing the Galactic SN rate of 3 SN/100yr to the radial dependence of
the surface density of either molecular clouds from Williams \& McKee (54) or
OB associations from McKee \& Williams (30), which are proportional to one
another, as we see in Table 1. Such a normalization gives a local supernova
rate at the Solar distance (8.5 kpc) of about 80 SN/Myr kpc$^2$ and a peak
rate at about 4 kpc of 250 SN/Myr kpc$^2$. From the calculated dependence
of the filling factor on supernova rate, we thus estimate the Galactic
radial dependence of the SB/HISM filling factor within $\mid z \mid <$300 pc,
as shown in Table 1. We see that the SB/HISM is expected to fill most
($\sim$ 90\%) of the ISM within $\mid z \mid <$300 pc from somewhere inside
of 4 kpc out to roughly the Solar distance of 8.5 kpc, decreasing thereafter
with Galactic radius to $\sim$ 50\% at 10 kpc and $\sim$ 10\% at 12 kpc where
the supernova rate is very low. As we show below, a SB/HISM filling factor
of $>$ 50\% can provide a cosmic ray injection composition in the SB/HISM
that is consistent with current estimates of the required cosmic ray source
composition. We note that one recent estimate by Ferriere (14) of the radial
dependence of the SB/HISM filling factor gives only 20\% locally, but this is
for a very low local supernova rate from a very steep assumed radial dependence
that is not consistent with the molecular cloud and OB association observations.

A local SB/HISM filling factor of $\sim$ 90\% would appear to be quite
consistent with observations within the local kpc, see Blaauw (2) Fig. 8,
which show that the Sun presently lies inside the $\sim$ 500 pc radius
superbubble produced by the $\sim$ 30 Myr Cas-Tau OB association, e.g. Olano
(37). This local superbubble is defined in the Galactic plane by a ring of
young OB associations know as Gould's Belt which have formed from the
ring of cooling gas swept up by the superbubble. The Cas-Tau association
inturn is part of a larger ($\sim$ 1 kpc radius) ring of OB associations,
including Cam-1, Aur-1, Gem-1 and Mon-2, formed by an older, now vanished
OB association.

\section{Cosmic Ray Acceleration in Superbubble/Hot ISM}
\label{acceleration.sec}

These hot, low density superbubbles are the hot phase of the ISM,
where shock acceleration of cosmic rays is expected, e.g. Axford (1),
to be ``most effective", because the energy losses of the accelerated
particles are greatly reduced and the supernova shocks do not suffer major
radiative losses, as they would in a denser medium. The rapid radiative
loss of supernova remnant energy in the average ISM sets in at a radius of
$\sim$20 pc, while the undiminished shock energy of nonradiative remnants
in the superbubble hot phase expand out to radii of $\sim$200 pc.
At full shock energy, supernovae in the low density SB/HISM expand to
$\sim$10$^3$ times the volume of those in the average ISM. Thus, the
supernova shocks in low density, but metal enriched SB/HISM process
a comparable masses of gas and for $z>2$ at least twice the metals as
those in the average ISM, contrary to the estimate of Ellison \& Meyer (12).

Also, since the energy of supernova shocks in the SB/HISM, unlike that
of shocks in the denser ISM, is not dissipated by radiation losses before
the shocks slow to sound speed, cosmic rays are accelerated in SB/HISM
primarily by low Mach number shocks. Such low Mach number (e.g. $<$4) shocks
can produce, e.g. Axford (1), the power-law index of $\sim$2.3 required for
the cosmic ray source spectrum, while the lower spectral indices ($\sim$2)
produced by high Mach number shocks in the denser ISM are not consistent
with the required source value.

The observed concentration of supernovae in the superbubble hot phase and
the much higher acceleration efficiency expected there clearly show that
the bulk of the cosmic rays must be accelerated in the SB/HISM. Such an
acceleration site also argues strongly against a first-ionization-potential
(FIP) injection bias, e.g. Meyer (32), which requires warm partially
ionized gas, not the highly ionized gas of the hot phase. Acceleration
in the SB/HISM further argues against a mass/charge (A/Q) dependent acceleration
model for the volatile elements, which Ellison \& Meyer (12) argue does
not work in highly ionized hot gas. As we have shown in Lingenfelter et
al. (25) and Lingenfelter \& Ramaty (24) and discuss further below, however,
sputtering and scattering of hot gas by high velocity refractory grains from
supernovae in the SB/HISM can provide a self-consistent cosmic ray injection
source for both refractory and volatile elements.

The transient acceleration of low energy ($<$100 MeV/nucleon) cosmic rays
(LECRs) in superbubbles has also been suggested, e.g. Bykov (5), as an
alternative source of Be production in the Galaxy. To account for the
measured Be/Fe evolution solely by LECRs, however, would require (Ramaty
et al. (41)) that there be as much or more energy in the LECRs as there
is the relativistic cosmic rays. Bykov (5) suggests that such LECRs
might be accelerated in supernova shocks during the early ($<$3 Myr)
stages of superbubble formation and that these LECRs are later further
accelerated to relativistic cosmic ray energies by the ensemble of supernova
shocks as the superbubble fully develops, e.g. Bykov \& Fleishman (6). But
since the energy in such LECRs persists for only a small fraction ($<$10\%)
of the age ($\sim$50 Myr) of the superbubble and then more energy is added as
the LECRs become relativistic cosmic rays which persist for most of
the age of the superbubble, such a model can not produce a time averaged
LECR energy comparable to that of the relativistic cosmic rays. Even if
the LECRs were not further accelerated to relativistic energies, comparable
total energy densities in LECRs and relativistic cosmic rays would require
that roughly half of the supernovae in superbubbles accelerate LECRS, but
$<$5\% of the superbubble supernovae occur during the first few Myr of
superbubble growth when condition favorable to LECR acceleration might be
expected (Bykov (5)).

\section{Expected Abundances of Refractory Cosmic Rays}
\label{abundances.sec}

We have shown in Lingenfelter et al. (25), Higdon et al. (17) and Lingenfelter
\& Ramaty (24) that the observed enrichment of the cosmic ray refractory
elements can be produced by the preferential acceleration in the SB/HISM
of suprathermal ions sputtered off high velocity (few 1000 km s$^{-1}$)
refractory grains, which formed as condensates in the expanding ejecta
of supernovae, e.g. Kozasa et al. (21) and Dwek (9).
The measured (Naya et al. (35)) broad width (5.4$\pm$1.4 keV) of the
Galactic 1.809 MeV line from the decay of long-lived (1.0x10$^6$ yr
mean life) $^{26}$Al, most likely produced in Type II supernovae, e.g.
Woosley \& Weaver (56), clearly suggests that refractory grains,
containing most of the live Galactic $^{26}$Al, are still moving at
velocities of $\sim$450 km s$^{-1}$ some 10$^6$ yrs after their formation,
and that the bulk of the grains are in low density superbubbles because the
grains would have been stopped much earlier in the much denser average ISM.
We also showed that only a very small fraction ($\sim$10$^{-4}$)
of the grains formed in a typical supernova need be accelerated
to account for the average injection of cosmic ray metals.

\begin{table}
\begin{tabular}{cccccccc}
\hline
&ISMGrains & ISMCores
\tablenote{ ISMGrains and ISMCores -- HST interstellar depletion determined
abundance from Savage \& Sembach (44).}
 & SNGrains\tablenote{
SNGrains -- Range of IMF averaged supernova ejecta mixes weighted with relative
SNII:SNIb:SNIa rates of 67-75\%:13-15\%:20-10\% from van den Berg \& Tammann
(50) and van den Berg \& McClure (49), except for O; for the SNII and SNIb
contributions, refractory O is assumed to be bound in MgSiO$_3$, Fe$_3$O$_4$,
Al$_2$O$_3$, CaO and NiO, and for (the very small) SNIa contribution, all the
produced O is assumed bound to Fe.}
 & SBGrains\tablenote{
SBGrains -- Modified SNGrains for 85\% of SNII and SNIb and 50\% of
SNIa in superbubbles plus ISM refractory grain ISMCores for a mean
superbubble metallicity range of 2--5 times that of ISM, as discussed in the
text.}
 & CRInject\tablenote{
 CRInject -- Galactic supernova averaged grain abundances for cosmic ray
injection, taking a mix of SBGrain abundances for supernova acceleration
in superbubbles and ISMCore grain abundances (without any supernova
enrichment) for supernova acceleration outside the superbubbles, weighted
by the relative swept-up metal masses and supernova rates, as discussed
in the text.}
 & CRSource\tablenote{
 CRSource -- elemental abundances from Engelmann et al. (13).}
 & Solar\tablenote{Solar system -- elemental abundances from Grevesse,
 Noels \& Sauval (15).} \\
\hline
C/Fe  & 690 & -?- & 210--510 &    --    &    --    & 422$\pm$14  &
1122$\pm$139\\
O/Fe  &1400 & 400 & 320--520 & 460--690 & 455--665 & 522$\pm$11  &
2344$\pm$414\\
Mg/Fe & 115 & 110 &  50--150 &  90--190 &  90--185 & 103$\pm$3   &
120$\pm$4\\
Al/Fe &  10 &  10 &   5--16  &   8--20  &   8--20  & 7.7$\pm$1.5 &
9.8$\pm$0.3\\
Si/Fe & 105 &  65 & 110--170 & 105--185 & 100--175 &  99$\pm$2   &
115$\pm$4\\
Ca/Fe &   6 &   6 &   4--8   &   5--9   &   5--9   & 6.0$\pm$0.9 &
7.1$\pm$0.2\\
Ni/Fe &   6 &   6 &   6--14  &   6--9   &   6--9   & 5.6$\pm$0.2 &
5.6$\pm$0.2\\
\hline
\end{tabular}
\caption{COSMIC RAY INJECTION ABUNDANCE RATIOS IN \%}
\label{tab:b}
\end{table}

The similarity of the cosmic ray source and solar abundance ratios
of refractory elements, mainly Mg, Al, Si, Ca relative to Fe, simply
reflects the fact that supernovae are the primary source of these
elements, e.g. Timmes et al. (47), and that the SB/HISM filling factor
is large where cosmic rays are accelerated, so that the bulk of
the Fe grains from the SNIa also contribute to the high velocity grain
population in the SB/HISM. In particular, since the Si, Mg, Al, and
other refractory elements are primarily produced in core-collapse
SNII/Ibc, while only about half of the Fe is made in them and the
other half is made in thermonuclear SNIa, a SB/HISM filling factor
of $\sim$ 90\% leads to differences of only $\sim$ 5\% between the
Si/Fe ratio in SB/HISM and the average Galactic production ratio,
which determines that in the well-mixed ISM. This is well within
the present uncertainties in the inferred cosmic ray source ratios
shown in Table 2, where we see from a much more detailed estimate
in Lingenfelter \& Ramaty (24) that the injection abundances expected
for cosmic ray acceleration predominantly in the SB/HISM is consistent
with the present cosmic ray source ratios of Engelmann et al. (13)
even for an assumed SB/HISM filling factor of only 50\%.
Similar small differences $<$ 10\% in relative abundances from a
SB/HISM filling factor of $\sim$ 90\% would be expected for those
s-process elements which appear to come primarily from the winds
of less massive stars.

The estimated mean refractory abundances in supernova grains (Table 2)
are based on the calculations by Woosley \& Weaver (56), Woosley, Langer
\& Weaver (55) and Nomoto et al. (36) of supernova ejecta abundances for
Types II, Ib and Ia, averaged over the initial mass function and
supernova rates of van den Berg \& Tammann (50) and van den Bergh \& McClure
(49), except that we assume the grain O abundance is limited to that bound
in Al$_{2}$O$_{3}$, MgSiO$_{3}$, Fe$_{3}$O$_{4}$, CaO and NiO. We also
show for comparison, the refractory abundances in the typical, older
icy interstellar grains (ISMGrains) and their refractory cores (ISMCores)
recently determined by HST observations, see Savage \& Sembach (44).
Here we see that the Si/Fe of 65\% in refractory cores of ISM grains,
which Meyer et al. (33) proposed as the cosmic ray source, is not
consistent with the required cosmic ray source value of 99$\pm$2\%.

\section{Expected Abundances of Volatile Cosmic Rays}
\label{abundances.sec}

In addition to the sputtering of refractory ions, the interactions of the
high velocity, supernova grains can also provide a simultaneous, self
consistent cosmic ray injection source of H, He and other volatiles.
Cesarsky \& Bibring (7) suggested that high velocity grains may temporarily
pick up by implantation volatile atoms from the gas through which they pass,
and their subsequent sputtering could provide a source of less enriched
suprathermal volatiles.  We suggest a much more direct injection process
for the volatiles. Since direct collisions of fast grains with ambient
gas atoms and ions are thought to be the primary means of grain momentum
loss, e.g. Ellison et al. (11) \S 2.3, we would expect that the
supernova grains should simply scatter ambient H, He and other volatile
atoms to the same suprathermal injection velocities as the grains and
their sputtered refractory products. Such a process would, in fact,
directly account for the measured cosmic ray abundance ratio by number
of the refractory (including C and ``bound" O) to volatile elements,
i.e. (C,O,Mg,Al,Si,Fe,etc)/(H,He,etc) = 0.010 of Engelmann et al. (13),
since Ellison et al. (11 \S 2.4) assume that roughly 0.5\%-1\% of
grain collisions with ambient gas atoms, predominantly scattering
volatile atoms, result in the sputtering of a refractory atom from
the grain surface, all of which come off with essentially the same
injection velocity. Moreover, because the geometric scattering cross
section increases with mass to the 2/3 power, such scattering should also
lead to a mass-dependent enrichment of heavier volatiles with respect to
H, as is observed in the cosmic rays, e.g. Meyer et al. (33), and
which Ellison \& Meyer (12) argue can not be accounted for by an
A/Z dependent acceleration bias in the hot ISM.

The composition of the grain-scattered suprathermal volatiles can be further
enriched by the fact that most of the supernova shocks will be interacting
with grains and gas in the supernova-ejecta and progenitor-wind enriched
superbubbles. Since the $^{22}$Ne/$^{20}$Ne ratio in the Wolf Rayet winds
of massive, supernova progenitors may exceed the solar system value by
more than two orders of magnitude, e.g. Maeder \& Meynet (27), grain-scattering
of such wind enriched could account for the high $^{22}$Ne/$^{20}$Ne observed
in the cosmic rays, e.g. Leske et al. (22). The existence of such
a Wolf Rayet signature in the cosmic rays also provides further evidence
for the acceleration of cosmic rays in the superbubble hot phase where the
bulk of the massive Wolf Rayet, supernova progenitors are also confined.

%\section{Acknowledgments}
This work was supported by NASA ATP and ACE/GI Programs.

\end{document}